\documentclass{ctr}

\usepackage{ctrfont}
\usepackage{natbib}

\usepackage{url}
\usepackage{amsmath}
\usepackage{amssymb}

\usepackage[dvips]{graphicx}
\usepackage{psfrag}
\usepackage{epsfig}

\usepackage{graphicx}
\usepackage{color}
\usepackage{psfrag}
\usepackage{subfigure}
\usepackage{hyperref}
\usepackage{amsmath,amsfonts,stmaryrd,amssymb,theorem}
\usepackage{xcolor,colortbl}
\usepackage[mathscr]{euscript}
\usepackage[ruled]{algorithm2e}
\usepackage{algorithmic}
\usepackage{changepage}
\usepackage{multirow}
\usepackage{rotating}
\usepackage{bm}
\usepackage{lineno}
\usepackage{enumitem}

\usepackage[nameinlink, capitalize]{cleveref}

\usepackage{tikz}

\usepackage{relsize}

\theoremstyle{plain}

\DeclareMathAlphabet{\mathpzc}{OT1}{pzc}{m}{it}

\usepackage{caption}
\crefname{figure}{Figure}{Figures}

\newcolumntype{C}[1]{>{\centering\let\newline\\\arraybackslash\hspace{0pt}}m{#1}}

\makeatletter
\newcommand\rlarrows{\mathop{\operator@font \rightleftarrows}\nolimits}
\makeatother
 \def\0vec{{\mbox{\boldmath$0$}}}

\title{Numerical methods to prevent pressure oscillations in transcritical flows}
\shorttitle{Numerical methods for transcritical flows}
\author{P. C. Ma, Y. Lv \and M. Ihme}
\shortauthor{Ma et al.}

\begin{document}


\maketitle

\section{Motivation and objectives} 

The accurate and robust simulation of transcritical real-fluid effects is crucial for many engineering applications, such as fuel injection in internal combustion engines, rocket engines and gas turbines. For example, in diesel engines, the liquid fuel is injected into the ambient gas at a pressure that exceeds its critical value, and the fuel jet will be heated to a supercritical temperature before combustion takes place. This process is often referred to as transcritical injection. The largest thermodynamic gradient in the transcritical regime occurs as the fluid undergoes a liquid-like to a gas-like transition when crossing the pseudo-boiling line \citep{yang2000modeling, oschwald2006injection, banuti2015crossing}. The complex processes during transcritical injection are still not well understood. Therefore, to provide insights into high-pressure combustion systems, accurate and robust numerical simulation tools are required for the characterization of supercritical and transcritical flows.

One obstacle in studying transcritical flows with variable thermodynamic properties is spurious pressure oscillations that are generated when a fully conservative scheme is adopted. This is similar to that of multi-component compressible ideal gas flow calculations \citep{abgrall2001computations}. However, due to the strong nonlinearities of the thermodynamic system, this issue is more severe in transcritical real-fluid flow predictions. This problem cannot be mitigated by introducing numerical dissipation using lower-order schemes or artificial viscosity, as pointed out in previous works \citep{kawai2015robust, schmitt2010large, terashima2012approach, ruiz2012unsteady, hickey2013large}. To overcome this issue, \cite{terashima2012approach} solved a transport equation for pressure instead of the total energy equation in their finite difference solver. This approach was inspired by the work of \cite{karni1994multicomponent} on multi-species calorically perfect gas, by which the pressure equilibrium can be maintained since pressure is explicitly solved for instead of derived from flow variables. \cite{schmitt2010large} derived a quasi-conservative scheme by connecting the artificial dissipation terms in the mass, momentum, and energy conservation equations and setting the pressure differential to zero. This procedure was later adopted by \cite{ruiz2012unsteady}. For calorically perfect gas flows, \cite{johnsen2012preventing} adopted a scheme in which an auxiliary advection equation for the specific heat ratio was solved. Although this method is suitable for well-defined interfacial flows with inert species, it is not applicable for complex transcritical and reacting flows in which the thermodynamic properties are dependent on temperature and species compositions. \cite{saurel1999multiphase} and \cite{saurel2009simple} developed methods for interfacial flows. However, these methods are typically limited to binary systems.

Another approach originally developed for a calorically perfect gas is the double-flux model. This method was proposed by \cite{abgrall2001computations}, extended by \cite{billet2003adaptive} for reacting flows, and later formulated for high-order schemes \citep{billet2011runge, houim2011low}. This quasi-conservative method has been reported to correctly predict shock speeds even for very strong shock waves \citep{abgrall2001computations}. In the present work, the double-flux method will be modified and extended to transcritical flows for general real-fluid state equations.

The remainder of this report has the following structure. \cref{sec:governing} introduces the governing equations and the description of thermodynamic relations. \cref{sec:numerics} discusses spurious pressure oscillations related to the solution of fully conservative formulations, and the development of the double-flux method for real-fluid transcritical flows. In \cref{sec:tests}, test cases are conducted to examine the performance and conservation properties of the double-flux model. The report finishes with conclusions in \cref{sec:conclusions}.


\section{Governing equations\label{sec:governing}}
The governing equations are the conservation of mass, momentum, total energy, and species, which take the following form
\begin{subequations}
    \label{eqn:governingEqn}
    \begin{align}
        \partial_t \rho +\nabla \cdot (\rho \boldsymbol{u}) &= 0\,,\\
        \partial_t (\rho \boldsymbol{u}) + \nabla \cdot (\rho \boldsymbol{u} \boldsymbol{u} + p \boldsymbol{I}) &= \nabla \cdot \boldsymbol{\tau} \,,\\
        \partial_t  (\rho e_t) + \nabla \cdot ((\rho e_t + p) \boldsymbol{u} ) &= \nabla \cdot (\boldsymbol{\tau} \cdot \boldsymbol{u}) -\nabla \cdot \boldsymbol{q} \,,\\
        \partial_t (\rho Y_k) + \nabla \cdot (\rho \boldsymbol{u} Y_k) &= \nabla \cdot (\rho D_k \nabla Y_k) \,,
    \end{align}
\end{subequations}
where $\rho$ is the density, $\boldsymbol{u}$ is the velocity vector, $p$ is the pressure, $e_t$ is the specific total energy, $Y_k$ is the mass fraction of species $k$, $D_k$ is the diffusion coefficient for species $k$, and $N_S$ is the number of species. Totally $N_S-1$ species transport equations are solved, and the system is closed by enforcing the total species conservation. The viscous stress tensor and heat flux are written as
\begin{subequations}
    \begin{align}
        &\boldsymbol{\tau} =  \mu \left[ \nabla \boldsymbol{u} + (\nabla \boldsymbol{u})^T \right] -\frac{2}{3}\mu (\nabla \cdot \boldsymbol{u}) \boldsymbol{I} \;,\\
        &\boldsymbol{q} = - \lambda \nabla T - \rho \sum_{k = 1}^{N_S} {h}_k D_k \nabla Y_k\;, \label{eqn:heatflux}
    \end{align}
\end{subequations}
where $T$ is the temperature, $\mu$ is the dynamic viscosity, $\lambda$ is the thermal conductivity, and ${h}_k$ is the partial enthalpy of species $k$. The specific total energy is related to the internal energy and the kinetic energy
\begin{equation}
    \rho e_t = \rho e + \frac{1}{2}\rho \boldsymbol{u} \cdot \boldsymbol{u}\,.
\end{equation}

The system of equations, \cref{eqn:governingEqn}, is closed with a state equation. The Peng-Robinson (PR) cubic equation of state (EoS)~\citep{peng1976new} is used in this study. It can be written as
\begin{equation}
    \label{EQ_PR_EQUATION_P}
    p = \frac{R T}{v - b} - \frac{a}{v^2+2bv-b^2}\,,
\end{equation}
where $R$ is the gas constant, $v$ is the specific volume, and the coefficients $a$ and $b$ are dependent on temperature and composition to account for effects of intermolecular forces. Extended corresponding states principle and pure fluid assumption for mixtures are adopted \citep{ely1981prediction, ely1983prediction}. The mixing rules, the procedures for evaluating thermodynamic quantities using PR-EoS, and the evaluation of transport properties in the transcritical regime can be found in \cite{hickey2013large} and \cite{ma2016entropy}.

A finite volume approach is utilized for the discretization of the system of equations. A Strang-splitting scheme \citep{strang1968construction} is applied to separate the convection and diffusion operators of the system. The numerical methods developed in this study focuses mainly on the hyperbolic operator. A strong stability-preserving third-order Runge-Kutta (SSP-RK3) scheme \citep{gottlieb2001strong} is used for the time integration of each operator.

\section{Double-flux method\label{sec:numerics}}
For a calorically perfect gas, the relation between internal energy and pressure can be written as
\begin{equation}
    e = \frac{pv}{\gamma-1} + e_0\;, \label{eqn:idealGasEnergy}
\end{equation}
where $\gamma$ is the specific heat ratio, and $e_0$ is the internal energy at a reference temperature. Without loss of generality, we can write the specific internal energy for a general fluid in the following form through an effective specific heat ratio,
\begin{equation}
    \label{eqn:internalEnergyDF}
    e = \frac{pv}{\gamma^*-1} + e^*_0 \,,
\end{equation}
where $\gamma^*$ and $e^*_0$ can be nonlinear functions of the thermodynamic states and may not be constant as for calorically perfect gases. To ensure that the characteristic speed of sound of the system is identical to the thermodynamic speed of sound, $\gamma^*$ is defined as
\begin{equation}
    \label{eqn:gammaS}
    \gamma^* = \frac{c^2 \rho}{p}\,,
\end{equation}
where $c$ is the speed of sound. Note that the definition of the effective specific heat ratio in this study is different from that in \cite{ma2014supercritical}.

The source of spurious pressure oscillations will be examined first before the introduction of an extension of the double-flux model for transcritical flows. Consider using \cref{eqn:internalEnergyDF} as the EoS, the Euler system of \cref{eqn:governingEqn} can be solved for a contact interface problem where pressure and velocity are constant at time step $t^n$. Using a first-order Godunov scheme with upwinding flux for one time step, following the analysis of \cite{billet2003adaptive}, we have the following expression for the change in pressure
\begin{equation}
    \label{eqn:DFPrinciple}
    \begin{split}
        \rho_j^{n+1} \delta e^*_{0,j} + p_j^n\delta\left(\frac{1}{\gamma^*_j-1}\right) + \frac{1}{\gamma^{*,n+1}_j-1}\delta p_j \\
       = -\sigma u_j^n \left[\rho_{j-1}^n \Delta e^{*,n}_0 + p_j^n\Delta\left(\frac{1}{\gamma^{*,n}-1}\right) \right] \,,    
    \end{split}
\end{equation}
where $\delta(\cdot) = (\cdot)^{n+1} - (\cdot)^n$ represents the temporal variation, $\Delta(\cdot) = (\cdot)_{j} - (\cdot)_{j-1}$ is the spatial variation, and $\sigma = \frac{\Delta t}{\Delta x}$. It can be seen that as long as $\gamma^*$ and $e^*_0$ are not constant, or that the internal energy $\rho e$ is not linear in both pressure and density, a pressure equilibrium cannot be maintained across the interface. This is true when a fully conservative scheme is used for the Euler system of \cref{eqn:governingEqn}. From \cref{eqn:DFPrinciple}, it can be seen that oscillations in pressure are related to the spatiotemporal discontinuity in $1/(\gamma^* - 1)$ and $e^*_0$.

\begin{figure}
 \centering
 \subfigure[Compressibility factor]{
    \includegraphics[width=0.452\columnwidth,clip=]{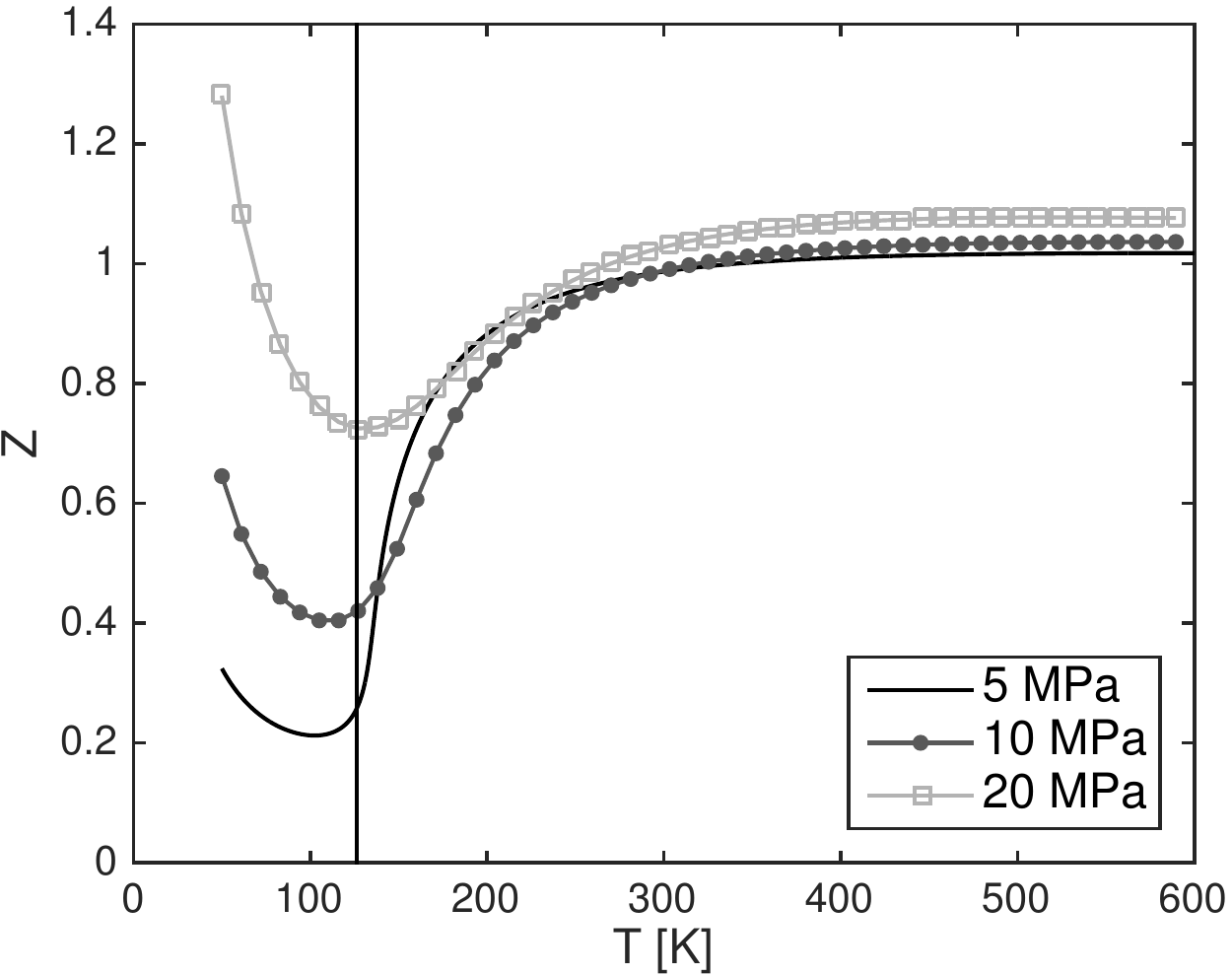}
  \label{fig:subfig1} 
 }   
 \subfigure[Effective specific heat ratio]{
   \includegraphics[width=0.498\columnwidth,clip=]{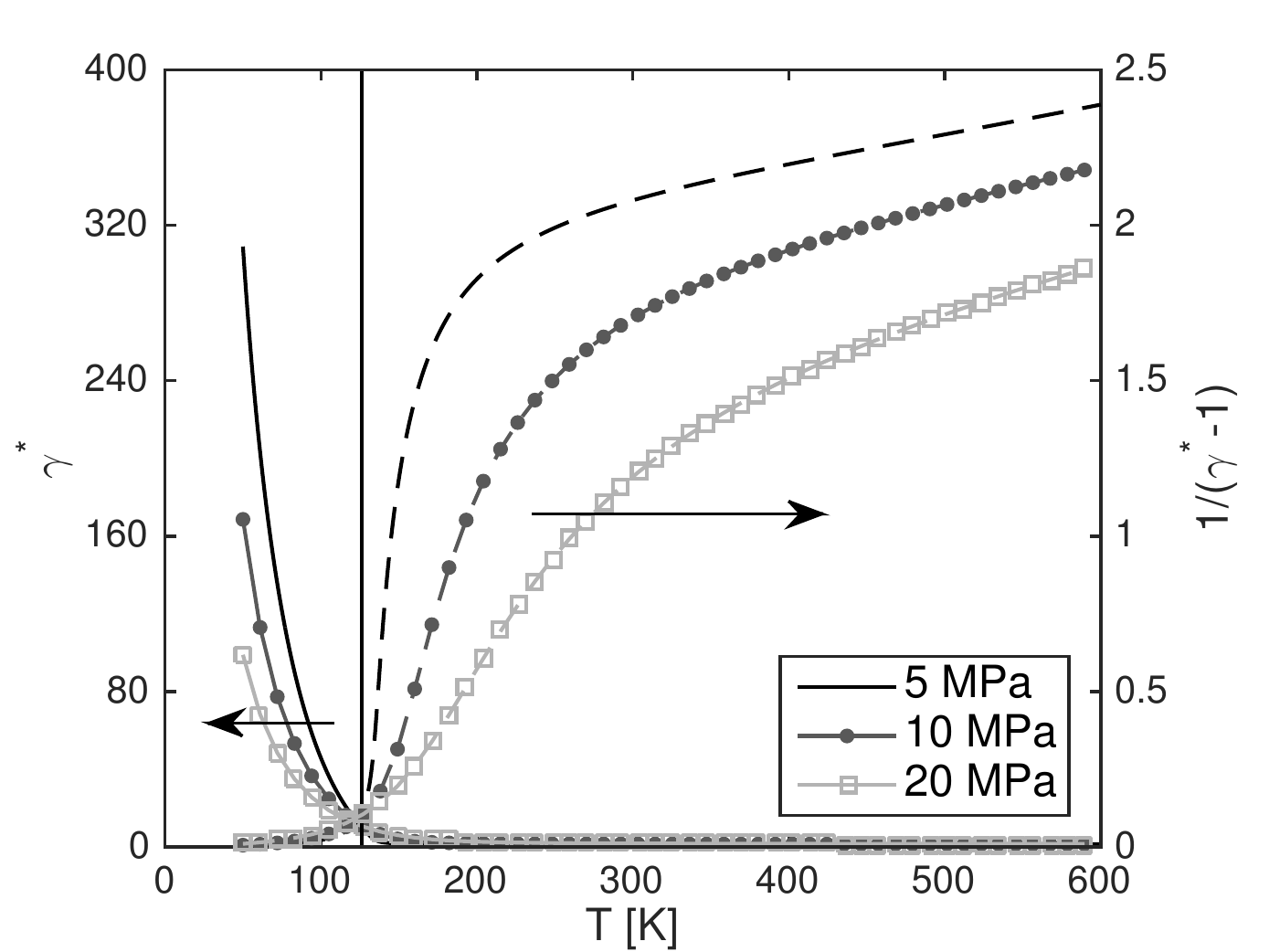}
  \label{fig:gammaSb} 
 }
 \caption{\label{fig:gammaS} Compressibility factor (a) and effective specific heat ratio (b) for nitrogen at various pressures. Vertical line indicates the critical temperature for nitrogen, which is 126.2 K. Solid lines in (b) are for $\gamma^*$ and dashed lines are for $1/(\gamma^*-1)$.}
\end{figure}

The nonlinearity between the coupling of internal energy, density and pressure could result from different situations, e.g., multi-fluid cases in which $\gamma^*$ is a nonlinear function of mass fractions, reacting flow cases in which $\gamma^*$ is a nonlinear function of temperature, and cases in which general EoS and compressibility are taken into account. For transcritical simulations, which often involve multi-species and chemically reacting flows with a general EoS, it can be seen that all the above situations are present. \Cref{fig:gammaS} shows the compressibility factor and the effective specific heat ratio of nitrogen at various pressures in the temperature range between 50 K and 600 K. The nonlinearity from the cubic EoS can be clearly seen. With increasing pressure, this nonlinearity is reduced since the difference between liquid and gas when crossing the pseudo-boiling region becomes smaller~\citep{oschwald2006injection, banuti2015crossing, Banuti2016sub}. The results for $\gamma^*$ also explain why pressure oscillations are substantially more severe for transcritical simulations than for conditions involving ideal gases. For example, at all pressures considered, $\gamma^*$ is $\mathcal{O}$(100) when the temperature is below the critical value; however, for gaseous conditions, the specific heat ratio is always $\mathcal{O}$(1). Moreover, across the narrow temperature range of the pseudo-boiling region, $\gamma^*$ changes dramatically, compared with the slowly increasing behavior in the higher-temperature region. The significant jump between liquid-like and gas-like fluid results in a large jump in $\gamma^*$, which causes significant spurious pressure oscillations in the transcritical regime.

The principle of the double-flux model is based on the results from \cref{eqn:DFPrinciple}. Since spurious pressure oscillations arise whenever $\gamma^*$ or $e_0^*$ is not constant, the main idea of the double-flux model is to locally freeze $\gamma^*$ and $e_0^*$ in both space and time during each time-step advancement \citep{abgrall2001computations, billet2003adaptive}.

For clarity, we consider first a one-dimensional case to derive the model, and this model is subsequently extended to multi-dimensions. The numerical Euler flux at the face $x_{j+\frac{1}{2}}$ can be evaluated as
\begin{equation}
    F^e_{j+\frac{1}{2}} = F^e(U^L_{j+\frac{1}{2}}, U^R_{j+\frac{1}{2}})\,,
\end{equation}
where $U^L_{j+\frac{1}{2}}$ and $U^R_{j+\frac{1}{2}}$ are the left and right reconstructed states at the face $x_{j+\frac{1}{2}}$, computed from the corresponding stencils. The first step in the double-flux model is to freeze $\gamma^*$ and $e_0^*$ for each cell in a stencil. During the reconstruction process for the face at $x_{j+\frac{1}{2}}$, whenever the total energy in the stencil is needed, the value is computed from the frozen values of $\gamma^*$ and $e_0^*$ through
\begin{equation}
    \label{eqn:DFEnergy}
    (\rho e_t)^n_l = \frac{p^n_l}{\gamma^{*,n}_j - 1} + \rho^n_l e^{*,n}_{0,j} + \frac{1}{2} \rho^n_l \boldsymbol{u}^n_l \cdot \boldsymbol{u}^n_l \,,
\end{equation}
in which $l$ is the index for the cells in the stencil. Note that when primitive variables are used for reconstruction, this step is applied only when the flux at the face is computed from primitive variables. If a central flux or a Riemann flux is used, the total energy of the left and right states is computed from \cref{eqn:DFEnergy}. From this relation, it can be seen that the numerical flux for the energy at face $x_{j+\frac{1}{2}}$ is different for cell $x_j$ and $x_{j+1}$ since different frozen values are used for $\gamma^*$ and $e_0^*$.

After the conservative variables are updated, the primitive variables for each cell are updated using the frozen values of $\gamma^*$ and $e_0^*$. Specifically, the pressure is updated using the following expression,
\begin{equation}
    \label{eqn:DFPressure}
    p_j^{n+1} = (\gamma^{*,n}_j-1)\left[(\rho e_t)^{n+1}_j - \rho^{n+1}_j e^{*,n}_{0,j} - \frac{1}{2} \rho^{n+1}_j \boldsymbol{u}^{n+1}_j \cdot \boldsymbol{u}^{n+1}_j \right]\,.
\end{equation}
This step ensures that $\gamma^*$ and $e_0^*$ are frozen in time.

The double-flux model has been implemented in an unstructured finite volume code, CharLES$^X$. Numerical tests will be conducted in \cref{sec:tests} to evaluate the performance and conservation error associated with the currently developed numerical scheme.

\section{Test cases\label{sec:tests}}

\subsection{Advection of the nitrogen interface}\label{sec:1DN2}
A one-dimensional advection configuration is selected to evaluate the performance of the proposed numerical schemes. Note that the Euler system is solved for all test cases in this subsection, which enables a direct comparison with the analytical solution.

The test case involves nitrogen as a working fluid. The pressure is set to 5 MPa, which is above the critical pressure of nitrogen. The computational domain is $x \in [0, 1]$~m and a uniform mesh is utilized. Periodic boundary conditions are applied. Two types of initial conditions are considered. One case involves a sharp jump of density, and the other case considers a smooth density profile. For the case with the sharp density jump, the initial conditions are
\begin{equation}
    \rho = \left\{
    \begin{array}{ll}
      \rho_\text{max}\,, \phantom{\rho_\text{min}} 0.25 < x < 0.75  \\
      \rho_\text{min}\,, \phantom{\rho_\text{max}} \text{otherwise}
    \end{array}
  \right. \,.
\end{equation}
For the case with the smooth density profile, a harmonic wave is given for the density with the same maximum and minimum as for the case with a sharp jump initial condition,
\begin{equation}
    \rho = \frac{\rho_\text{min}+\rho_\text{max}}{2} + \frac{\rho_\text{max}-\rho_\text{min}}{2} \text{sin} (2 \pi x)\,. 
\end{equation}
The two states of nitrogen correspond to temperature and density at $T_\text{min} = 100$~K, $\rho_\text{max} = 793.1$~kg/m$^3$ and $T_\text{max} = 300$~K, $\rho_\text{min} = 56.9$~kg/m$^3$, respectively. For all computations, the CFL number is set to a value of 0.8. The advection velocity is 100 m/s for all cases and the simulation is run for one period, corresponding to a physical time of 0.01 s.

A hybrid scheme \citep{johnsen2010assessment, hickey2013large} is utilized to deal with the large density gradients present in the test cases. A relative solution (RS) sensor \citep{khalighi2011unstructured, ma2015discontinuous} is used in the present study. For each control volume (CV), the RS sensor determines whether the face density value from high-order non-dissipative reconstruction, $\rho_f$, and the density value at the CV center, $\rho_{cv}$, differ by more than some fraction of the density at the CV center, which can be expressed as
\begin{equation}
    \left| \frac{\rho_{f} - \rho_{cv}}{\rho_{cv}} \right| > \zeta\,,
\end{equation}
where $\zeta$ is a user-input parameter. 

\begin{figure}
    \centering
    \includegraphics[width=14.4cm,clip=]{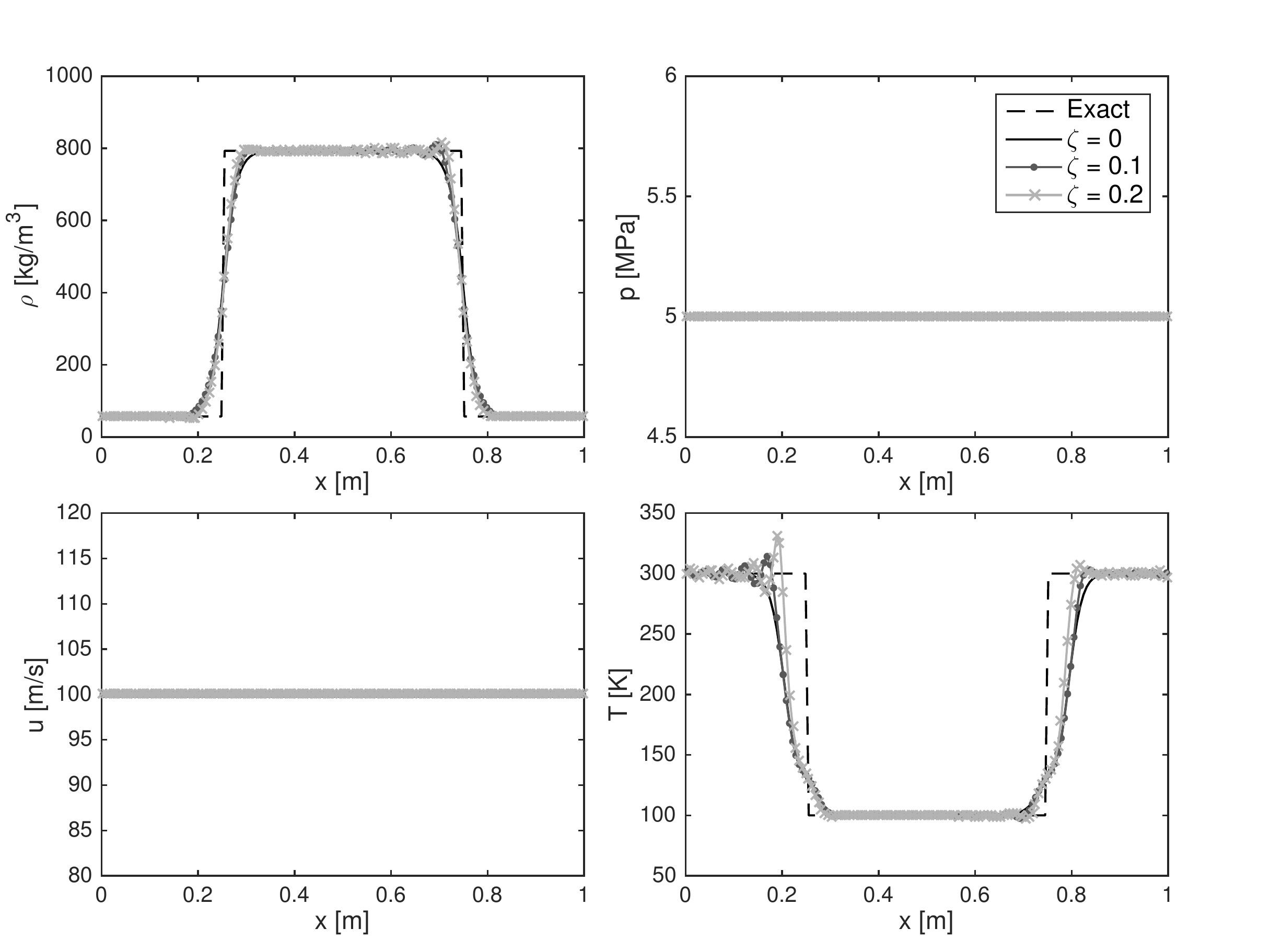}
    \caption{\label{fig:1DN2_RS} Results of density, pressure, velocity, and temperature for the one-dimensional N$_2$ advection case with a sharp jump as the initial condition after one period at $t = $~0.01~s. A uniform mesh is used with mesh size $N$ = 150.}
\end{figure}

\Cref{fig:1DN2_RS} shows results for density, pressure, velocity, and temperature for the case with a sharp jump as the initial condition. The exact solution for this case is the advection of the initial condition with constant pressure and velocity. To examine the behavior of the hybrid scheme, simulations with different threshold values of the RS sensor are performed. As can be seen in the pressure profiles in \cref{fig:1DN2_RS}, the present numerical scheme with the double-flux model preserves the pressure and velocity equilibrium without generating spurious oscillations. In contrast, the fully conservative scheme will fail the simulation at the first sub-iteration of the SPP-RK3 time advancement due to the negative pressure that is generated from the strong oscillations as a result of the large discontinuity in $\gamma^*$ and $e_0^*$. Refining the grid will not solve the problem unless the initial condition is smooth and can be fully resolved. The effect of using different sensor values in the hybrid scheme is clearly seen in the density and temperature profiles in \cref{fig:1DN2_RS}. Primitive variables are used for the reconstruction in the ENO scheme for all test cases in this study, and no significant difference was observed using reconstruction from conservative variables. For the case with $\zeta$ = 0, the ENO scheme is applied everywhere in the computational domain, and the results do not show any visible oscillations. As the sensor value is relaxed, the solution for density and temperature becomes less diffusive and the interface becomes sharper, but small oscillations in density become apparent. Due to the non-linear thermodynamics, oscillations that are introduced by numerical instabilities are subsequently magnified in the temperature profiles.

\begin{figure}
    \centering
    \subfigure[Sharp initial condition]{
        \includegraphics[width=0.48\columnwidth,clip=]{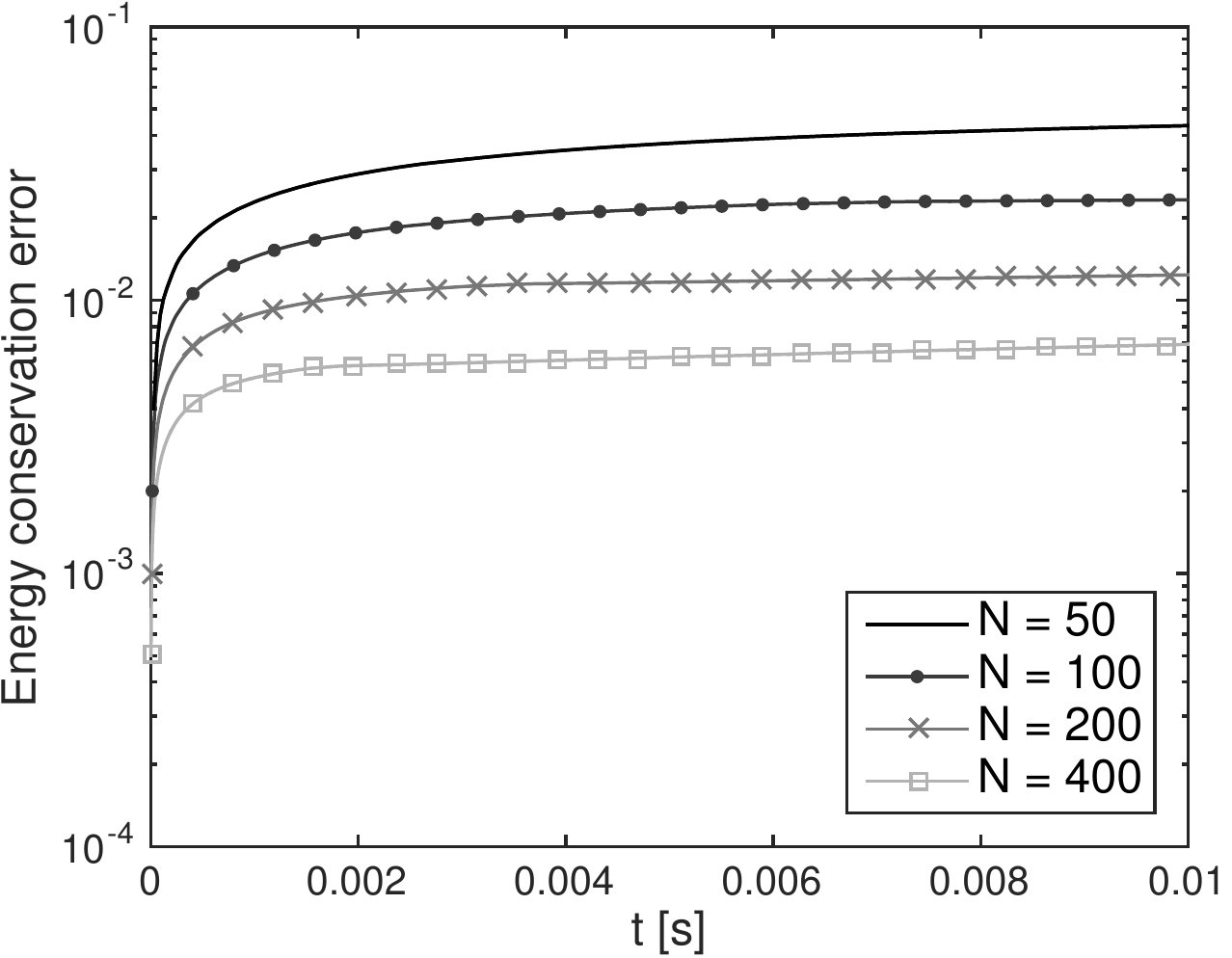}
        \label{fig:subfig1} 
    }
    \subfigure[Smooth initial condition]{
        \includegraphics[width=0.48\columnwidth,clip=]{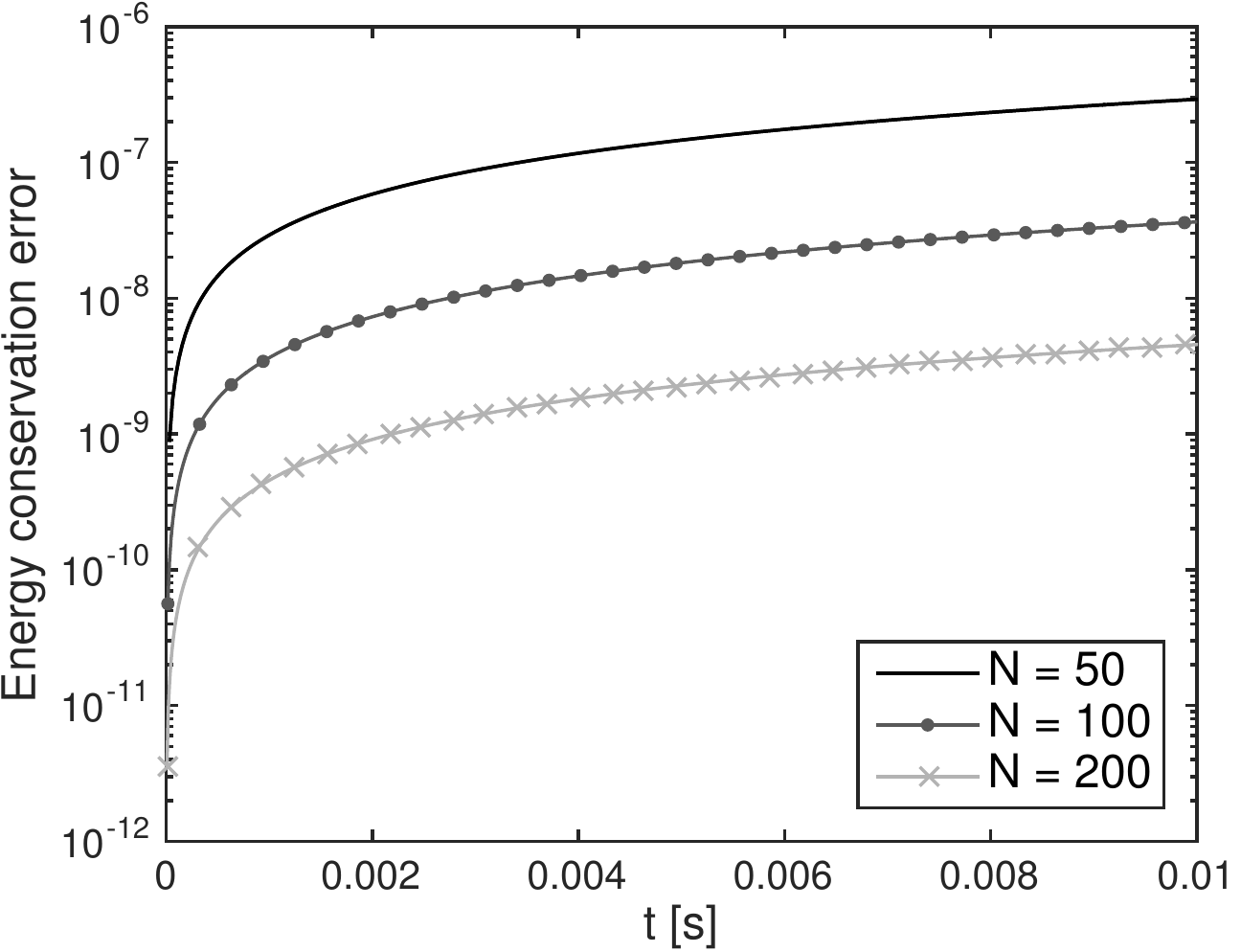}
        \label{fig:subfig1} 
    }
    \caption{\label{fig:1DN2_conservation} Energy conservation error for the one-dimensional N$_2$ advection case for (a) sharp jump initial condition and (b) smooth initial profile; $\zeta$ = 0.2.}
\end{figure}

To study the conservation error introduced by the double-flux model, the evolution of the energy conservation error is studied for both cases with discontinuous and smooth initial conditions. The energy conservation error is defined as
\begin{equation}
    \epsilon = \left|\frac{\int_\Omega (\rho e_t)(t) dx - \int_\Omega (\rho e_t)(0) dx}{ \int_\Omega(\rho e_t)(0) dx}\right|\,,
\end{equation}
which is the relative error of the total energy with respect to initial conditions and $\Omega$ represents the computation domain. Results with $\zeta$ = 0.2 are shown in \cref{fig:1DN2_conservation}. Simulations with other RS values produce similar behaviors and are therefore omitted. For all cases, the conservation error initially increases rapidly before approaching a plateau. This is similar to the findings from previous studies by \cite{abgrall2001computations}. For the case with a discontinuous initial condition, the conservation error at the end of one period shows a first-order mesh convergence. Whereas for the case with a smooth density profile, a much faster convergence can be observed. This is due to the fact that for smooth solutions, the profiles for $\gamma^*$ and $e_0^*$ are also smooth and as the mesh is refined, the jump in thermodynamic relations is reduced in addition to the decrease in mesh size $\Delta x$. Mass and momentum are conserved by the current numerical scheme to the level of machine error. Overall, the conservation error in total energy of the current numerical scheme with the double-flux model is acceptable and converges to zero with mesh refinement.

\subsection{Diffusion of n-dodecane into nitrogen}
To study the performance of the double-flux model on diffusion-dominated problems, a one-dimensional problem of liquid {\it n}-dodecane diffusing into a gaseous nitrogen environment is considered.

The operating conditions for this test case correspond to the Spray A conditions~\citep{pickett2011engine}. The pressure is 6 MPa. The {\it n}-dodecane drop has a temperature of 363~K and the hot nitrogen environment has a temperature of 900~K. The computational domain is $x \in [-10, 10]$~$\mu$m and a uniform mesh is utilized. Initially, the pressure is set to be constant and the velocity is set to be zero. Initial conditions for temperature and mass fraction are set based on the adiabatic mixing profiles. Periodic boundary conditions are applied corresponding to an isochoric problem. For all computations, the CFL number is set to a value of 1.0. An ENO scheme is applied for all the faces ($\zeta = 0$). The enthalpy diffusion term in the heat flux (\cref{eqn:heatflux}) is neglected. The simulation is run for 2$\times 10^{-6}$~s. Note that this case is designed to test the conservation performance of the double-flux model in comparison with the fully conservative scheme on diffusion-dominated problems.

\begin{figure}
    \centering
    \includegraphics[width=14.4cm,clip=]{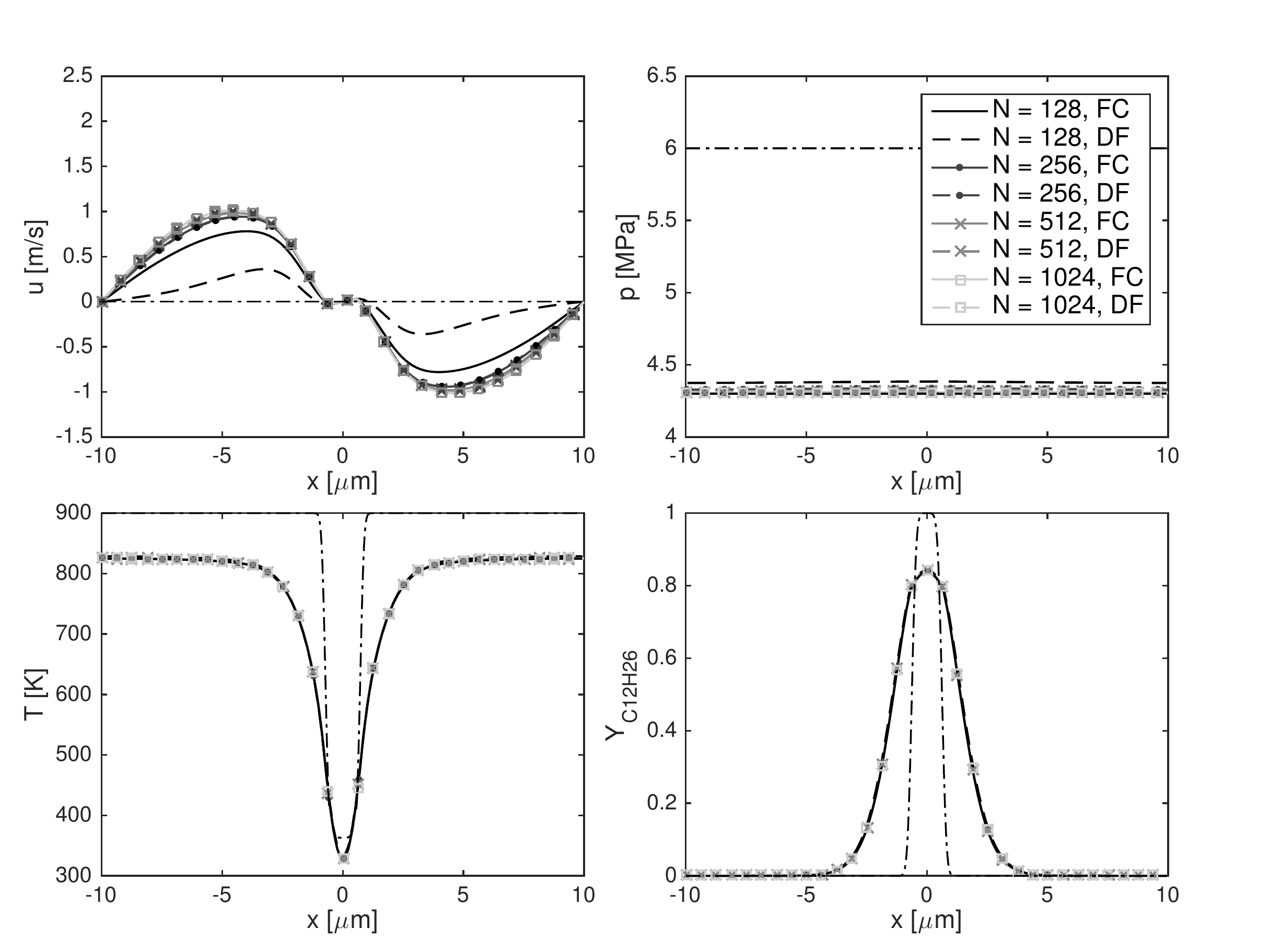}
    \caption{\label{fig:1DNC12H26} Results of velocity, pressure, temperature, and {\it n}-dodecane mass fraction for the one-dimensional diffusion case at $t$~=~2$\times 10^{-6}$~s. Results from fully conservative (FC) and double-flux (DF) method are shown. Dash-dotted line denotes initial conditions.}
\end{figure}

\Cref{fig:1DNC12H26} shows results for velocity, pressure, temperature, and mass fraction of {\it n}-dodecane at the end of the simulations. For each grid resolution, the simulation is conducted with both a fully conservative scheme and the double-flux model to examine the performance. A small velocity is induced by the diffusion processes, and the pressure decreases rapidly due to the nonlinearity of the real-fluid state relations. Since the initial conditions are smooth and the problem is diffusion dominated, the spurious oscillations in pressure and velocity are not expected when the fully conservative scheme is utilized. It can be seen in \cref{fig:1DN2_RS} that with grid refinement, for both the fully conservative scheme and the double-flux scheme, temperature and mass fraction profiles start to converge with 256 grid points. However, the convergence in velocity and pressure profiles is slower; even with 1024 grid points, the velocity profile is still not fully converged and a small difference from 512 grid points can be observed. For the conservation properties, it can be seen that results from the double-flux model become almost identical to those from the fully conservative scheme except for the case with the coarsest grid, demonstrating the convergence of the conservation error. This can be expected from the results in the previous section, as the small conservation error coming from the convection part using the double-flux model is converging to zero with grid refinement.

\subsection{Cryogenic nitrogen injection}
The test cases in this subsection examine the performance of the present numerical scheme in three dimensions in the context of LES. To this end, the cryogenic nitrogen injection case of \cite{mayer2003raman} is considered. In the experiment, a cryogenic supercritical nitrogen jet is injected into an ambient nitrogen reservoir at 298 K and 3.97 MPa. Case 3 in the experiment is simulated in this study, corresponding to injection conditions of 126.9 K, 440 kg/m$^3$, and an injection speed of 4.9 m/s. The injector diameter, $d$, is 2.2 mm and the dimensions of the domain are the same as those in the experiment. Cryogenic nitrogen is injected with a plug-flow profile and no smoothing or turbulence profile is used. Isothermal no-slip conditions are applied at the cylinder face, and adiabatic no-slip conditions are applied at the cylinder wall. The pressure is specified at the outlet boundary condition. For the simulations in this subsection, the governing equations, \cref{eqn:governingEqn}, are Favre-filtered and the Vreman subgrid-scale model~\citep{vreman2004eddy} is used as the closure for turbulence. A grid convergence study is conducted using three different meshes with mesh sizes of 4, 8, and 17 million cells, respectively. The unstructured mesh is generated from a base hexahedral mesh and then adapted and refined in the region of interest. The minimum resolution of the mesh is in the region near the injector to resolve the large density gradients across the shear layer.

\begin{figure}
    \centering
    \includegraphics[width=0.8\columnwidth,clip=]{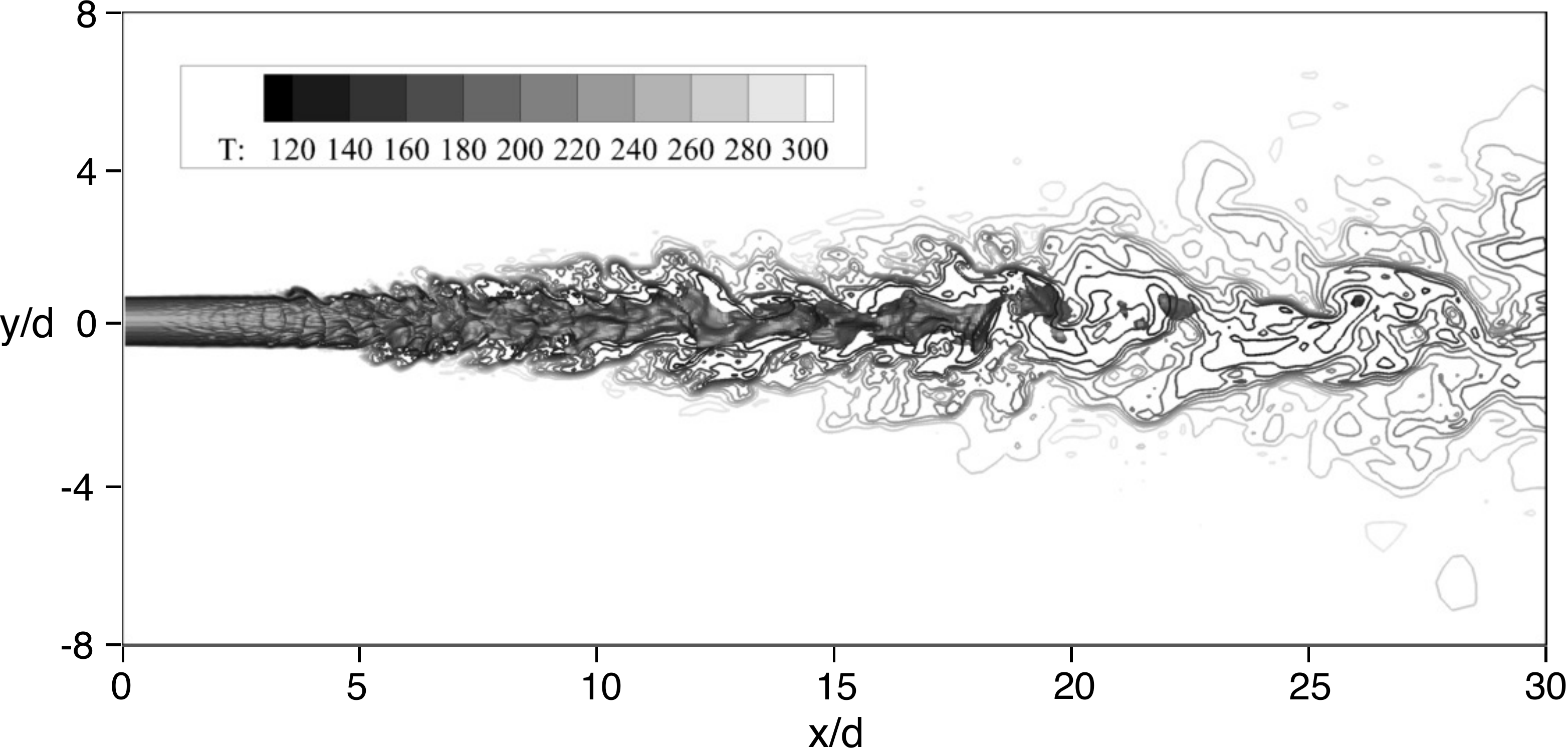}
    \caption{\label{fig:Mayer_results} Instantaneous flow field in the cryogenic N$_2$ injection case. The density iso-surface with values equal to 243 kg/m$^3$ is plotted on top of the temperature contours (in Kelvin) at the center-plane. Result from medium mesh is shown.}
\end{figure}

\begin{figure}
    \centering
    \includegraphics[width=7.8cm,clip=]{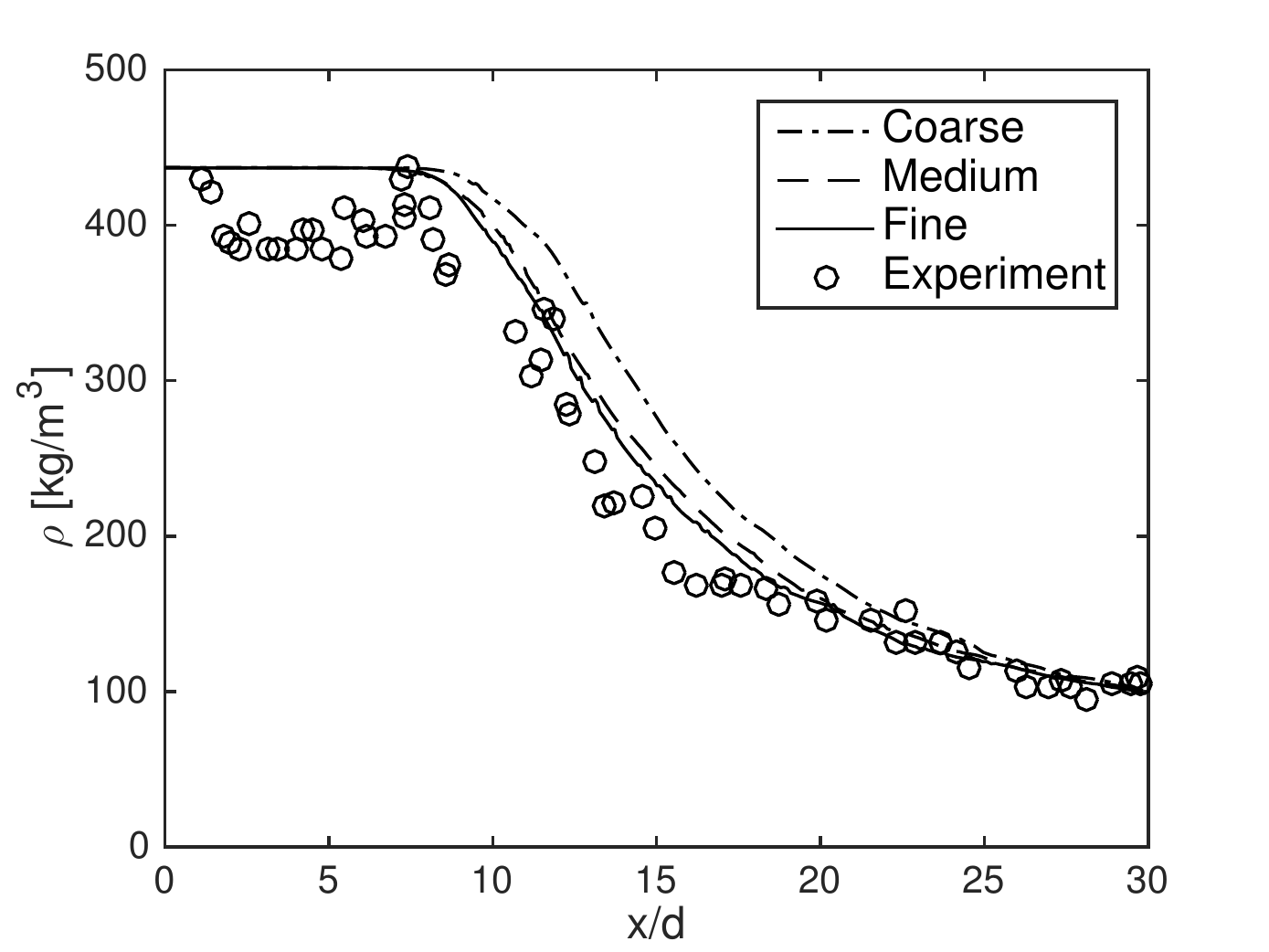}
    \caption{\label{fig:Mayer_mean} Mean centerline density profiles of the cryogenic N$_2$ injection case in comparison with experimental data from \cite{mayer2003raman}.}
\end{figure}

\Cref{fig:Mayer_results} shows instantaneous results of the simulation. The density iso-surface of values equal to 243 kg/m$^3$ is shown and the temperature contours at the center-plane are also plotted. The density values are chosen to be the average of the injection and ambient densities. No spurious pressure or velocity oscillations were seen during the simulation, even on the coarse mesh. The simulation results are averaged over four flow-through times to obtain statistics, and one flow-through time corresponds to the time for the jet to convect 30$d$ at the injection velocity. \Cref{fig:Mayer_mean} shows mean centerline density profiles using different meshes. As can be seen in \cref{fig:Mayer_mean}, the results from medium and fine meshes have almost identical mean density profiles, demonstrating mesh convergence of the simulations. In comparison with the experimental measurements from \cite{mayer2003raman}, the simulation results in the current study capture the behavior of the cryogenic nitrogen jet in the experiment. The capability of the present numerical scheme for LES on unstructured meshes for transcritical real-fluid simulations is demonstrated.

\section{Conclusions\label{sec:conclusions}} 
A double-flux model is extended in the context of finite volume schemes to prevent spurious pressure oscillations in transcritical flows. Due to the large jumps in thermodynamic properties that arise from nonlinearities in the real-fluid EoS, the pressure oscillations are more severe for transcritical flows than for ideal gas conditions and could cause the divergence of the simulation. A double-flux model is formulated for transcritical simulations by introducing an effective specific heat ratio based on the speed of sound to eliminate spurious pressure oscillations. One-dimensional test cases are conducted at different operating conditions relevant to practical applications. No spurious oscillations in pressure and velocity are generated for convection-dominated test cases with sharp initial conditions which cannot be handled by fully conservative schemes. The conservation errors generated by the double-flux formulation are acceptable and converge to zero rapidly for smooth solutions. The multi-dimensional cryogenic injection case demonstrates the capability of the currently developed numerical scheme for the simulations of real applications.

\section*{Acknowledgments} 

This investigation was funded by NASA with award numbers NNX14CM43P and NNM13AA11G. The authors would like to thank Prof. Stefan Hickel and Jan Matheis for helpful discussions.


 





\begin{thebibliography}{30}
\expandafter\ifx\csname natexlab\endcsname\relax\def\natexlab#1{#1}\fi

\bibitem[Abgrall \& Karni(2001)]{abgrall2001computations}
{\sc Abgrall, R. \& Karni, S.} 2001 Computations of compressible multifluids.
  {\em J. Comput. Phys.\/} {\bf 169}~(2), 594--623.

\bibitem[Banuti(2015)]{banuti2015crossing}
{\sc Banuti, D.~T.} 2015 Crossing the {W}idom-line--{S}upercritical
  pseudo-boiling. {\em J. Supercrit. Fluids\/} {\bf 98}, 12--16.

\bibitem[Banuti {\em et~al.\/}(2016)Banuti, Ma, Hickey \& Ihme]{Banuti2016sub}
{\sc Banuti, D.~T., Ma, P.~C., Hickey, J.-P. \& Ihme, M.} 2016 Sub- or
  supercritical? {A} flamelet analysis of high pressure rocket propellant
  injection. In {\em 52nd AIAA/SAE/ASEE Joint Propulsion Conference\/}, pp.
  2016--4789. Salt Lake City, UT.

\bibitem[Billet \& Abgrall(2003)]{billet2003adaptive}
{\sc Billet, G. \& Abgrall, R.} 2003 An adaptive shock-capturing algorithm for
  solving unsteady reactive flows. {\em Comput. Fluids\/} {\bf 32}~(10),
  1473--1495.

\bibitem[Billet \& Ryan(2011)]{billet2011runge}
{\sc Billet, G. \& Ryan, J.} 2011 A {R}unge--{K}utta discontinuous {G}alerkin
  approach to solve reactive flows: the hyperbolic operator. {\em J. Comput.
  Phys.\/} {\bf 230}~(4), 1064--1083.

\bibitem[Ely \& Hanley(1981)]{ely1981prediction}
{\sc Ely, J.~F. \& Hanley, H.} 1981 Prediction of transport properties. 1.
  {Viscosity} of fluids and mixtures. {\em Ind. Eng. Chem. Res.\/} {\bf
  20}~(4), 323--332.

\bibitem[Ely \& Hanley(1983)]{ely1983prediction}
{\sc Ely, J.~F. \& Hanley, H.} 1983 Prediction of transport properties. 2.
  {Thermal} conductivity of pure fluids and mixtures. {\em Ind. Eng. Chem.
  Res.\/} {\bf 22}~(1), 90--97.

\bibitem[Gottlieb {\em et~al.\/}(2001)Gottlieb, Shu \&
  Tadmor]{gottlieb2001strong}
{\sc Gottlieb, S., Shu, C.-W. \& Tadmor, E.} 2001 Strong stability-preserving
  high-order time discretization methods. {\em SIAM Rev.\/} {\bf 43}~(1),
  89--112.

\bibitem[Hickey {\em et~al.\/}(2013)Hickey, Ma, Ihme \&
  Thakur]{hickey2013large}
{\sc Hickey, J.-P., Ma, P.~C., Ihme, M. \& Thakur, S.} 2013 Large eddy
  simulation of shear coaxial rocket injector: {Real} fluid effects. In {\em
  49th AIAA/ASME/SAE/ASEE Joint Propulsion Conference\/}, pp. 2013--4071. San
  Jose, CA.

\bibitem[Houim \& Kuo(2011)]{houim2011low}
{\sc Houim, R.~W. \& Kuo, K.~K.} 2011 A low-dissipation and time-accurate
  method for compressible multi-component flow with variable specific heat
  ratios. {\em J. Comput. Phys.\/} {\bf 230}~(23), 8527--8553.

\bibitem[Johnsen \& Ham(2012)]{johnsen2012preventing}
{\sc Johnsen, E. \& Ham, F.} 2012 Preventing numerical errors generated by
  interface-capturing schemes in compressible multi-material flows. {\em J.
  Comput. Phys.\/} {\bf 231}~(17), 5705--5717.

\bibitem[Johnsen {\em et~al.\/}(2010)Johnsen, Larsson, Bhagatwala, Cabot, Moin,
  Olson, Rawat, Shankar, Sj{\"o}green, Yee {\em
  et~al.\/}]{johnsen2010assessment}
{\sc Johnsen, E., Larsson, J., Bhagatwala, A.~V., Cabot, W.~H., Moin, P.,
  Olson, B.~J., Rawat, P.~S., Shankar, S.~K., Sj{\"o}green, B., Yee, H. {\em
  et~al.\/}} 2010 Assessment of high-resolution methods for numerical
  simulations of compressible turbulence with shock waves. {\em J. Comput.
  Phys.\/} {\bf 229}~(4), 1213--1237.

\bibitem[Karni(1994)]{karni1994multicomponent}
{\sc Karni, S.} 1994 Multicomponent flow calculations by a consistent primitive
  algorithm. {\em J. Comput. Phys.\/} {\bf 112}~(1), 31--43.

\bibitem[Kawai {\em et~al.\/}(2015)Kawai, Terashima \&
  Negishi]{kawai2015robust}
{\sc Kawai, S., Terashima, H. \& Negishi, H.} 2015 A robust and accurate
  numerical method for transcritical turbulent flows at supercritical pressure
  with an arbitrary equation of state. {\em J. Comput. Phys.\/} {\bf 300},
  116--135.

\bibitem[Khalighi {\em et~al.\/}(2011)Khalighi, Nichols, Lele, Ham \&
  Moin]{khalighi2011unstructured}
{\sc Khalighi, Y., Nichols, J.~W., Lele, S., Ham, F. \& Moin, P.} 2011
  Unstructured large eddy simulation for prediction of noise issued from
  turbulent jets in various configurations. In {\em 17th AIAA/CEAS
  Aeroacoustics Conference\/}, pp. 2011--2886. Portland, Oregon.

\bibitem[Ma {\em et~al.\/}(2014)Ma, Bravo \& Ihme]{ma2014supercritical}
{\sc Ma, P.~C., Bravo, L. \& Ihme, M.} 2014 Supercritical and transcritical
  real-fluid mixing in diesel engine applications. {\em Proceedings of the
  Summer Program, Center for Turbulence Research, Stanford University\/} pp.
  99--108.

\bibitem[Ma {\em et~al.\/}(2015)Ma, Lv \& Ihme]{ma2015discontinuous}
{\sc Ma, P.~C., Lv, Y. \& Ihme, M.} 2015 Discontinuous {Galerkin} scheme for
  turbulent flow simulations. {\em Annual Research Briefs, Center for
  Turbulence Research, Stanford University\/} pp. 225--236.

\bibitem[Ma {\em et~al.\/}(2017)Ma, Lv \& Ihme]{ma2016entropy}
{\sc Ma, P.~C., Lv, Y. \& Ihme, M.} 2017 An entropy-stable hybrid scheme for
  simulations of transcritical real-fluid flows. {\em J. Comput. Phys.\/} {\bf
  340}, 330--357.

\bibitem[Mayer {\em et~al.\/}(2003)Mayer, Telaar, Branam, Schneider \&
  Hussong]{mayer2003raman}
{\sc Mayer, W., Telaar, J., Branam, R., Schneider, G. \& Hussong, J.} 2003
  Raman measurements of cryogenic injection at supercritical pressure. {\em
  Heat Mass Transfer.\/} {\bf 39}~(8-9), 709--719.

\bibitem[Oschwald {\em et~al.\/}(2006)Oschwald, Smith, Branam, Hussong, Schik,
  Chehroudi \& Talley]{oschwald2006injection}
{\sc Oschwald, M., Smith, J., Branam, R., Hussong, J., Schik, A., Chehroudi, B.
  \& Talley, D.} 2006 Injection of fluids into supercritical environments. {\em
  Combust. Sci. Technol.\/} {\bf 178}~(1-3), 49--100.

\bibitem[Peng \& Robinson(1976)]{peng1976new}
{\sc Peng, D.-Y. \& Robinson, D.~B.} 1976 A new two-constant equation of state.
  {\em Ind. Eng. Chem. Res.\/} {\bf 15}~(1), 59--64.

\bibitem[Pickett \& Bruneaux(2011)]{pickett2011engine}
{\sc Pickett, L. \& Bruneaux, G.} 2011 Engine combustion network. {\em
  Combustion Research Facility, Sandia National Laboratories, Livermore, CA.
  (http://www.sandia.gov/ECN)\/} .

\bibitem[Ruiz(2012)]{ruiz2012unsteady}
{\sc Ruiz, A.} 2012 Unsteady numerical simulations of transcritical turbulent
  combustion in liquid rocket engines. PhD thesis, Institut Nationale
  Polytechnique de Toulouse, France.

\bibitem[Saurel \& Abgrall(1999)]{saurel1999multiphase}
{\sc Saurel, R. \& Abgrall, R.} 1999 A multiphase {Godunov} method for
  compressible multifluid and multiphase flows. {\em J. Comput. Phys.\/} {\bf
  150}~(2), 425--467.

\bibitem[Saurel {\em et~al.\/}(2009)Saurel, Petitpas \&
  Berry]{saurel2009simple}
{\sc Saurel, R., Petitpas, F. \& Berry, R.~A.} 2009 Simple and efficient
  relaxation methods for interfaces separating compressible fluids, cavitating
  flows and shocks in multiphase mixtures. {\em J. Comput. Phys.\/} {\bf
  228}~(5), 1678--1712.

\bibitem[Schmitt {\em et~al.\/}(2010)Schmitt, Selle, Ruiz \&
  Cuenot]{schmitt2010large}
{\sc Schmitt, T., Selle, L., Ruiz, A. \& Cuenot, B.} 2010 Large-eddy simulation
  of supercritical-pressure round jets. {\em AIAA J.\/} {\bf 48}~(9),
  2133--2144.

\bibitem[Strang(1968)]{strang1968construction}
{\sc Strang, G.} 1968 On the construction and comparison of difference schemes.
  {\em SIAM J. Num. Anal.\/} {\bf 5}~(3), 506--517.

\bibitem[Terashima \& Koshi(2012)]{terashima2012approach}
{\sc Terashima, H. \& Koshi, M.} 2012 Approach for simulating gas--liquid-like
  flows under supercritical pressures using a high-order central differencing
  scheme. {\em J. Comput. Phys.\/} {\bf 231}~(20), 6907--6923.

\bibitem[Vreman(2004)]{vreman2004eddy}
{\sc Vreman, A.} 2004 An eddy-viscosity subgrid-scale model for turbulent shear
  flow: Algebraic theory and applications. {\em Phys. Fluids.\/} {\bf 16}~(10),
  3670--3681.

\bibitem[Yang(2000)]{yang2000modeling}
{\sc Yang, V.} 2000 Modeling of supercritical vaporization, mixing, and
  combustion processes in liquid-fueled propulsion systems. {\em Proc. Combust.
  Inst.\/} {\bf 28}~(1), 925--942.

\end{thebibliography}

\end{document}